\documentclass{emulateapj} 
\usepackage{amsmath,amssymb,amsfonts} 
\usepackage{natbib}
\def\s{{\rm\,s}}

\def\m{{\rm\,m}} 
\def\km{{\rm\,km}}

\def\OmegaK{\Omega_{\rm K}}

\def\taus{\tau_{\rm s}}

\def\rhog{\rho_{\rm g}}
\def\rhoref{\rho_{\rm g,ref}}

\def\rhod{\rho_{\rm d}}
\def\rhodzero{\rho_{\rm d,ref}}

\def\Sigmad{\Sigma_{\rm d}}
\def\Sigmag{\Sigma_{\rm g}}

\def\cs{c_{\rm s}}

\def\lesssim{\mathrel{\hbox{\rlap{\hbox{\lower4pt\hbox{$\sim$}}}\hbox{$<$}}}}
\def\gtrsim{\mathrel{\hbox{\rlap{\hbox{\lower4pt\hbox{$\sim$}}}\hbox{$>$}}}}

\def\vmax{v_{\rm max}}
\def\Arms{A_{\rm rms}}
\newcommand{\p}{\partial}
\newcommand{\w}{\widetilde}
\begin{document}

\title{Forming Planetesimals by Gravitational Instability: \\ I. The Role of the
  Richardson Number in Triggering the Kelvin-Helmholtz Instability}

\author{Aaron T. Lee\altaffilmark{1}, Eugene Chiang\altaffilmark{1,2},
  Xylar Asay-Davis\altaffilmark{3},  and Joseph Barranco\altaffilmark{4}}

\altaffiltext{1}{Department of Astronomy, University of California,
    Berkeley, CA 94720}
\altaffiltext{2}{Department of Earth and Planetary Science, University of California,
    Berkeley, CA 94720}
\altaffiltext{3}{Center for Nonlinear Studies, Los Alamos National Laboratory, Los Alamos, NM 87545}    
\altaffiltext{4}{Department of Physics and Astronomy, San Francisco State University,
    San Francisco, CA 94132}
\email{a.t.lee@berkeley.edu}
\begin{abstract}
  Gravitational instability (GI) of a dust-rich layer at the midplane
  of a gaseous circumstellar disk is one proposed mechanism to form
  planetesimals, the building blocks of rocky planets and gas giant
  cores.  Self-gravity competes against the Kelvin-Helmholtz
  instability (KHI): gradients in dust content drive a vertical shear
  which risks overturning the dusty subdisk and forestalling GI.  To
  understand the conditions under which the disk can resist the KHI,
  we perform three-dimensional simulations of stratified subdisks in
  the limit that dust particles are small and aerodynamically well
  coupled to gas, thereby screening out the streaming instability and
  isolating the KHI. Each subdisk is assumed to have a vertical
  density profile given by a spatially constant Richardson number
  $Ri$. We vary $Ri$ and the midplane dust-to-gas ratio $\mu_0$ and
  find that the critical Richardson number dividing KH-unstable from
  KH-stable flows is not unique; rather $Ri_{\rm crit}$ grows nearly
  linearly with $\mu_0$ for $\mu_0$ = 0.3--10.  Plausibly a linear
  dependence arises for $\mu_0 \ll 1$ because in this regime the
  radial Kepler shear replaces vertical buoyancy as the dominant
  stabilizing influence.  Why this dependence should persist at $\mu_0
  > 1$ is a new puzzle.  The bulk (height-integrated) metallicity is
  uniquely determined by $Ri$ and $\mu_0$.  Only for disks of bulk
  solar metallicity is $Ri_{\rm crit} \approx 0.2$, close to the
  classical value. Our empirical stability boundary is such that a
  dusty sublayer can gravitationally fragment and presumably spawn
  planetesimals if embedded within a solar metallicity gas disk
  $\sim$$4 \times$ more massive than the minimum-mass solar nebula; or
  a minimum-mass disk having $\sim$$3\times$ solar metallicity; or
  some intermediate combination of these two possibilities. Gravitational instability seems
  possible without resorting to the streaming instability or to
  turbulent concentration of particles.
\end{abstract}
\section{INTRODUCTION}
\label{sec:introduction}
In the most venerable scenario for forming planetesimals, dust
particles in circumstellar gas disks are imagined to settle
vertically into thin sublayers (``subdisks'') sufficiently dense to undergo
gravitational instability (\citealt{safronov69};
\citealt{goldreichward73}; for a review of this and other ways in
which planetesimals may form, see Chiang \& Youdin 2010, hereafter
CY10). Along with this longstanding hope comes a longstanding fear
that dust remains lofted up by turbulence.  Even if we suppose that
certain regions of the disk are devoid of magnetized turbulence
because they are too poorly ionized to sustain magnetic activity
\citep{gammie96,baigoodman09}, the dusty sublayer is 
susceptible to a Kelvin-Helmholtz shearing instability
\citep[KHI;][]{Weiden80}.\footnote{\citet{goldreichward73} also recognized
that the sublayer would be shear-unstable, but unlike \citet{Weiden80},
overlooked the possibility that the KHI may forestall gravitational
instability.}\vspace{0.4in}

\subsection{Basic Estimates}
The KHI arises because dust-rich
gas at the midplane rotates at a different speed from dust-poor gas at
altitude.  The background radial pressure gradient $\partial
P/\partial r$ causes dust-free gas at disk radius $r$ to rotate at the
slightly non-Keplerian rate
\begin{equation}
\Omega_{\rm F} = \Omega_{\rm K} (1-\eta) \label{eqn:omegaf}
\end{equation}
where $\OmegaK$ is the Kepler angular frequency,
\begin{equation}
\label{eqn:eta}
\eta = 
 \frac{ -(1/\rho_{\rm g}) \partial P / \partial r }{2 \Omega_{\rm K}^2r} \approx 
 8 \times 10^{-4} \left( \frac{r}{\rm AU} \right)^{4/7} 
\end{equation}
is a dimensionless measure of centrifugal support by pressure, and
$\rhog$ is the density of gas
\citep[e.g.,][]{nakagawaetal86, cuzzietal93}. The numerical evaluation is based on the minimum-mass
solar nebula derived by CY10.  Unlike dust-free gas, dust-rich gas is
loaded by the extra inertia of solids and must rotate at more nearly
the full Keplerian rate to remain in centrifugal balance.  Variations
in the dust-to-gas ratio $\rhod/\rhog$ with height $z$ result in a
vertical shear $\partial v_{\phi}/\partial z$ from which free energy
is available to overturn the dust layer.

The
shearing rate across a layer of thickness $\Delta z$ is given to order
of magnitude by
\begin{eqnarray}
\left| \frac{\partial v_\phi}{\partial z} \right| \sim \frac{\Delta v_{\phi}}{\Delta z} & = & \frac{1}{\Delta z} \frac{\mu_0}{1+\mu_0} \eta \OmegaK r\nonumber \\
 & \approx & \frac{25}{\Delta z} \frac{\mu_0}{1+\mu_0} \left( \frac{r}{\rm AU} \right)^{1/14} \m \s^{-1} \label{eqn:verticalshear}
\end{eqnarray}
where $\rhod/\rhog = \mu_0$ at the midplane and $\rhod/\rhog \ll 1$ at
altitude (for more details see CY10, or
\S\S\ref{ssec:wcfe}--\ref{ssec:initialconditions} of this paper).  For
$\mu_0 \gg 1$ the velocity difference $\Delta v_\phi$ saturates at a
speed $\eta \OmegaK r \sim 25 (r/{\rm AU})^{1/14} \m \s^{-1}$, well
below the gas sound speed $c_{\rm s} \sim 1 \km \s^{-1}$. That the flow
is highly subsonic motivates what simulation methods we employ in our study.

We might expect the flow to be stabilized if the
Brunt-V\"ais\"al\"a frequency
\begin{eqnarray}
\omega_{\rm Brunt} & = &\left( \frac{-g}{\rho} \frac{\partial \rho}{\partial z} \right)^{1/2}  \nonumber \\
 & \sim & \left( \frac{\mu_0}{1+\mu_0} \right)^{1/2} \OmegaK \label{eqn:brunt}
\end{eqnarray}
of buoyant vertical oscillations is much larger than the
vertical shearing rate.
For the order-of-magnitude evaluation in (\ref{eqn:brunt})
we approximate the
vertical gravitational acceleration $g$ as the vertical component of stellar gravity $-\OmegaK^2 \Delta z$ (no
self-gravity), and the density gradient $\rho^{-1} \partial \rho
/ \partial z \sim (\rhod+\rhog)^{-1} \Delta (\rhod+\rhog)/ \Delta z
\sim (\rhod+\rhog)^{-1} \Delta \rhod / \Delta z$. The last
approximation relies in part on the dust density $\rhod$ changing over
a lengthscale $\Delta z$ much shorter than the gas scale height.
Both $|\partial v_{\phi}/\partial z|$ and $\omega_{\rm Brunt}$ shrink
as $\mu_0$ decreases.\footnote{But not indefinitely. In the limit
  $\mu_0 \rightarrow 0$, the vertical shearing and Brunt frequencies
  reach minima set by pressure and temperature gradients in gas \citep[see, e.g.,][]{knoblochspruit85}. The limit $\mu_0 \rightarrow 0$ is
  not relevant for our study and not captured by either
  (\ref{eqn:verticalshear}) or (\ref{eqn:brunt}).}

For two-dimensional, heterogeneous, unmagnetized flow, a necessary but not
sufficient condition for instability is given by the Richardson
number:
\begin{equation}
\label{eqn:richardson}
	Ri \equiv \frac{-(g/\rho)(d\rho/dz)}{(dv_\phi/dz)^2} < 1/4 \,\,\,{\rm is\,\,\, necessary\,\,\, for \,\,\, instability}
\end{equation}
(\citealt{miles61}; see the textbook by \citealt{drazinreid04}).
The Richardson number is simply the
square of the ratio of the stabilizing Brunt frequency
(\ref{eqn:brunt}) to the destabilizing vertical shearing frequency
(\ref{eqn:verticalshear}).  The critical value of 1/4 arises formally 
but can also be derived heuristically by energy arguments \citep[e.g.,][]{chandra81}. The Richardson criterion does not formally apply to our dusty
subdisk, which represents a three-dimensional flow: the KHI couples
vertical motions to azimuthal motions, while the Coriolis force
couples azimuthal motions to radial motions. 
(For how the Richardson criterion may not apply to magnetized
flows, see \citealt{lecoanetetal10}.)
Nevertheless we may hope
the Richardson number is useful as a guide, as previous works
have assumed \citep{sekiya98,youdinshu02,youdinchiang04}.

In this spirit let us use the Richardson criterion to estimate
the thickness of a marginally KH-unstable dust layer.
Substitution of (\ref{eqn:verticalshear})
and (\ref{eqn:brunt}) into (\ref{eqn:richardson}) reveals
that\footnote{This order-of-magnitude expression for the dust layer thickness, and the related equation (\ref{eqn:verticalshear}) which approximates the vertical shear, are each smaller than their counterparts given by \citet[][page 499, first full paragraph]{youdinshu02} by a factor of $(1+\mu)$. This is because \citet{youdinshu02} evaluate quantities deep inside the layer, within a density cusp at the midplane, whereas we are interested in quantities averaged across the entire layer. The difference does not change either our conclusions or theirs.}
\begin{equation}
\Delta z \approx \left( \frac{\mu_0}{1+\mu_0} \right)^{1/2} Ri^{1/2} \eta r \,.
\label{eqn:deltaz}
\end{equation}
Since the gas scale height $H_{\rm g} = \cs/\OmegaK$ and $\eta \sim
(H_{\rm g}/r)^2$, equation (\ref{eqn:deltaz}) indicates that for
$\mu_0 > 1$ the marginally
unstable dust sublayer is $\sim$$Ri^{1/2} H_{\rm g}/r \sim 0.02
Ri^{1/2}$ times as thin as the gas disk in which it is immersed.
Those KH-unstable modes that disrupt the layer should have
azimuthal wavelengths---and by extension radial wavelengths,
because the Kepler shear turns azimuthal modes into radial ones---that
are comparable to $\Delta z$. Shorter wavelength
modes cannot overturn the layer, while longer wavelength modes grow
too slowly \citep{gomezostriker05}.

How does disk rotation affect the development of the KHI?
In a linear analysis,
\citet{ishitsusekiya03} highlight the role played by the Keplerian
shear, characterized by the strain rate
\begin{equation}
\left| \frac{\partial \OmegaK}{\partial \ln r} \right| = \frac{3\OmegaK}{2} \,,
\label{eqn:keplershear}
\end{equation}
in limiting the growth of KH-unstable modes.
The radial shear is implicated because azimuthal motions excited by the KHI are converted
to radial motions by the Coriolis force; moreover, the non-axisymmetric
pattern excited by the KHI is wound up, i.e., stretched azimuthally
by the radial shear. The Kepler rate
$|\partial \OmegaK/\partial \ln r|$ is at least as large as
$\omega_{\rm Brunt}$, and can dominate the latter when $\mu_0$ is
small. This suggests that $Ri$ does not capture all the relevant
dynamics---a concern already clear on formal grounds.
In this paper we address this concern head-on, using fully 3D
numerical simulations to assess the role
of the Richardson number in governing the stability of the dust layer.

\subsection{Our Study in Relation to Previous Numerical Simulations}
Three-dimensional shearing box simulations of the KHI in dusty
subdisks, performed in the limits that dust is perfectly coupled to
gas and disk self-gravity is negligible, demonstrate the importance of
the Kepler shear. Compared to rigidly rotating disks
\citep{gomezostriker05,jhk06},
radially shearing disks are far more stable \citep{chiang08,barranco09}.
The relevance of $Ri$, or lack thereof, may be assessed by simulating flows with initially spatially constant $Ri$
\citep{sekiya98,youdinshu02}, and varying $Ri$ from run to run to see whether dust layers turn over.
\citet[][hereafter C08]{chiang08} found that when $\mu_0 > 1$, 
dust layers for which $Ri < 0.1$ overturn, while those for which $Ri > 0.1$ do
not.  In retrospect, we might have anticipated this result, that the critical 
value $Ri_{\rm crit}$ dividing stable from unstable runs
lies near the canonical value of
1/4, at least for $\mu_0 > 1$, because in this regime of parameter
space all the frequencies of the problem are comparable to each
other: $|\partial v_\phi/\partial z| \sim \omega_{\rm Brunt} \sim
|\partial \OmegaK/\partial \ln r| \sim \OmegaK$ when $\mu_0 > 1$ and
$Ri \approx 0.1$--1. But other simulations of C08 also make clear that $Ri$ does
not alone determine stability under all circumstances. For $\mu_0
\approx 0.2$--0.4, 
$Ri_{\rm crit}$ was discovered to drop substantially
to $\sim$0.02 (see his runs S9--S12). \citet{chiang08} speculated that
the baroclinic nature of the flow may be responsible \citep{knoblochspruit85, knoblochspruit86}, but no details were given.

In addition to being left unexplained, the findings of C08 
require verification.  Parameter space was
too sparsely sampled to discern trends with confidence. Concerns
about numerics---e.g., biases introduced by box sizes that were too small, 
resolutions too coarse, and runs terminated too early---also linger.
At least one numerical
artifact marred the simulations of C08: the KHI manifested
first at the ``co-rotation'' radius where the mean azimuthal flow
speed was zero (see his figure 8). But in a shearing box, by Galilean
invariance, there should be no special radius. It was suspected, but not
confirmed, that errors of interpolation associated with the grid-based advection
scheme used by C08 artificially suppressed the KHI away from co-rotation.

For the problem at hand, the spectral code developed by \citet{barrancomarcus06} and modified by 
\citet[][hereafter B09]{barranco09} to treat
mixtures of dust and gas is a superior tool to the grid-based ZEUS code
utilized by C08. Working in Fourier space rather than configuration
space, the simulations of the KHI by B09 did not betray the co-rotation
artifact mentioned above. Spectral methods, often used to model local
(WKB) dynamics, are appropriate here because the structures of
interest in the subdisk have dimensions tiny compared to the disk
radius (by at least a factor $Ri^{1/2} \eta$ according to equation
\ref{eqn:deltaz}) and even the gas scale height.  At the same
computational expense, spectral algorithms typically achieve greater
effective spatial resolution than their grid-based counterparts
\citep{barrancomarcus06}. Another advantage enjoyed by
the B09 code is that it employs the anelastic approximation, which is
designed to treat subsonic flows such as ours.  Having filtered away
sound waves, anelastic codes are free to take timesteps set by how
long it takes fluid to advect across a grid cell (which themselves
move at the local orbital velocity in a shearing coordinate system).
By contrast, codes such as ZEUS take mincing steps limited by the time for
sound waves to cross a grid cell. The latter constraint is the usual
Courant condition for numerically solving problems in compressible
fluid dynamics. It was unnecessarily applied by C08 to a practically
incompressible flow.

In this paper we bring all the advantages of the spectral, anelastic,
shearing box code of B09 to bear on the problems originally addressed
by C08.  We assess numerically the stability of flows characterized by
constant Richardson number $Ri$, systematically mapping out the
stability boundary in the parameter space of $Ri$, midplane
dust-to-gas ratio $\mu_0$, and bulk metallicity $\Sigmad/\Sigmag$ (the
height-integrated surface density ratio of dust to gas). Though our
simulations may still be underresolved, we rule out box size as a
major influence on our results.  We offer some new insight into why
$Ri$ is not a sufficient predictor of stability. And in the restricted
context of our constant $Ri$ flows, we assess the conditions necessary
for the midplane to become dense enough to trigger gravitational
instability on a dynamical time.

\subsection{The Perfect Coupling Approximation vs.~The Streaming Instability~vs.~Turbulent Concentration Between Eddies}\label{ssec:versus}
Following C08 and B09, we continue to work in the limit that dust is
perfectly coupled to gas, i.e., in the limit that particles are small
enough that their frictional stopping times $t_{\rm stop}$
in gas can be neglected in comparison to the dynamical time $\OmegaK^{-1}$. The
perfect coupling approximation allows us to screen out the streaming
instability which relies on a finite stopping time and which is most
powerful when particles are marginally coupled, i.e., when $\taus
\equiv \OmegaK t_{\rm stop} \sim 0.1$--1 \citep{youdingoodman05}.
Numerical simulations have shown that when an order-unity fraction of the
disk's solids is in particles having $\taus = 0.1$--1, the streaming
instability clumps them strongly and paves the way for gravitational
instability \citep[e.g.,][]{johansenetal07,johansenetal09}.
The particle sizes corresponding to $\taus = 1$ depend on
the properties of the background gas disk, as well as on
the particle's shape and internal density; under typical assumptions,
marginally coupled particles are decimeter to meter-sized.

It remains debatable whether a substantial fraction of a disk's solid
mass is in marginally coupled particles at the time of
planetesimal formation, as current proposals relying on the streaming
instability assume.  Particle size and shape distributions are not
well constrained by observations (though see, e.g.,
\citealt{wilneretal05}, who showed that centimeter-wavelength fluxes
from a few T Tauri stars are consistent with having been emitted by
predominantly centimeter-sized particles). Measuring $\taus$ in disks
also requires knowing the gas density, but direct measurements of the
gas density at disk midplanes do not exist.  Marginally coupled
particles---sometimes referred to as ``meter-sized boulders''---also
face the longstanding problem that they drift onto the central star
too quickly, within hundreds of years
from distances of a few
AU in a minimum-mass disk. \citet{johansenetal07} claimed to solve this problem by
agglomerating all the boulders into Ceres-mass planetesimals via the
streaming instability before they drifted inward. Their simulation
presumed, however, that all of the disk's solids began
boulder-sized. The concern we have is that even if particle-particle
sticking could grow boulders (and sticking is expected to stall
at centimeter sizes; \citealt{blumwurm08}; CY10), the disk's solids
may not be transformed into boulders all at once. Rather, 
marginally coupled bodies may initially comprise a minority population
on the extreme tail of the particle size distribution. Unless they can
transform themselves from a minority to a majority within the radial
drift timescale, they would be lost from the nebula by aerodynamic drag.

By focussing on the dynamics of the smallest, most well entrained
particles having $\taus \ll 1$, our work complements that which relies
on the streaming instability. 
We would argue further that the well
coupled limit is potentially more relevant for planet
formation.  If even the smallest particles having sizes $\ll$ cm can
undergo gravitational collapse to form kilometer-sized or larger
planetesimals, nature will have leapfrogged over the marginally
coupled regime, bypassing the complications and uncertainties described above.

Particle clumping is not restricted to marginally coupled particles
via the streaming instability. Small $\taus$ particles also clump
within the interstices of turbulent, high vorticity eddies \citep[][and references therein; for a review, see CY10]{maxey87,eatonfessler94, cuzzietal08}. This particle
concentration mechanism presumes some gas turbulence, which may be
present in the marginally KH-unstable state to which dust settles. Our
simulations cannot capture this phenomenon. However, on the scales of
interest to us, turbulent clumping might only be of minor
significance.  Particles of given $t_{\rm stop}$ are concentrated
preferentially by eddies that turn over on the same timescale. Thus
the degree of concentration depends sensitively on particle size and
the turbulent spectrum. At least in Kolmogorov turbulence, the
smallest eddies concentrate particles most strongly because they have
the greatest vorticity. The smallest eddies at the inner scale of
Kolmogorov turbulence have sizes $\ell_{\rm i} \sim \nu^{3/4}t_{\rm
  o}^{1/4}/\delta v_{\rm o}^{1/2}$, where $\nu$ is the molecular
kinematic viscosity, and $t_{\rm o}$ and $\delta v_{\rm o}$ are the
turnover time and velocity of the largest, outer scale eddy.  Given
$\delta v_{\rm o} \sim \eta \OmegaK r \sim 25 (r/{\rm AU})^{1/14}$
m/s, $t_{\rm o} \sim \OmegaK^{-1}$, and values of $\nu$ based on the
nebular model of CY10, we estimate that $\ell_{\rm i} \sim 10^3
(r/{\rm AU})^{127/56}$ cm. This is far smaller than the sublayer
thicknesses $\Delta z \sim 0.02 Ri^{1/2} H_{\rm g} \sim 2 \times
10^{9} (Ri/0.1)^{1/2} (r/{\rm AU})^{9/7}$ cm considered in this
paper. Moreover, the lifetimes of the particle clumps on a given eddy
length scale should roughly equal the eddy turnover times, which for
the smallest eddies are of order $t_{\rm i} \sim \sqrt{\nu t_{\rm
    o}}/\delta v_{\rm o} \sim 10^2 (r/{\rm AU})^{55/28}$ s.  We do not
expect such rapid fluctuations in particle density, occurring on such
small length scales, to affect significantly the evolution of the
slower, larger scale KHI. Turbulent clumping may only serve as a
source of noise on tiny scales. The possibility that
turbulent clumping could still be significant on larger scales
is still being investigated \citep{hogancuzzi07,cuzzietal08}.

The perfect coupling approximation prevents us from studying how
particles sediment out of gas into dusty sublayers, but it does not
stop us from identifying what kinds of sublayers are dynamically
stable to the KHI. A subdisk with a given density profile is either
dynamically stable or it is not, and we can run the B09 code for many
dynamical times (typically 60 or more) to decide the answer.  In a forthcoming
paper we will combine the B09 code with a settling algorithm that will
permit us to study how dust settles from arbitrary initial conditions,
freeing us from the assumption that the density profile derives from a
constant Richardson number.

\subsection{Organization of this Paper}
Our numerical methods, including our rationale for choosing
box sizes and resolutions, are described in \S\ref{sec:method}.
Results are presented in \S\ref{sec:results} and discussed
in \S\ref{sec:discussion}.


\section{METHODS}
\label{sec:method}

The equations solved by the B09 code are
rederived in \S\ref{ssec:wcfe}. Initial conditions for our simulations
are given in \S\ref{ssec:initialconditions}.
The code itself is briefly described in \S\ref{ssec:code}.
Our choices for box size and resolution are explained
in \S\ref{ssec:bsnr}.

\subsection{Equations}
\label{ssec:wcfe}

The equations we solve are identical to equations (12a--e) of B09.
We outline their derivation here, filling in steps skipped by B09,
adjusting the notation, and providing some clarifications. This section
may be skimmed on a first reading.

We begin with the equations for an ideal gas perfectly coupled to pressureless dust in an inertial frame:
\begin{eqnarray}
\frac{d \mathbf{v}}{dt} & = & -\nabla\Phi - \frac{\nabla P}{\rho_d + \rho_g}, \label{eqn:momentum}\\
\frac{d\rho_{\rm g}}{dt} & = & -\rho_g \nabla \cdot \mathbf{v}, \label{eqn:gascontinuity}\\
\frac{d(\rho_{\rm d}/\rho_{\rm g})}{dt} & = & 0, \label{eqn:noslip}\\
\rho_g C_{\rm V} \frac{dT}{dt} & = & -P \nabla \cdot \mathbf{v}, \label{eqn:adiabat} \\
P & = & \Re \rho_{\rm g} T, \label{eqn:idealeos}
\end{eqnarray}
\noindent 
where $d/dt$ is the convective derivative, $\rho_{\rm g(d)}$ is the
density of gas (dust), $P$ is the gas pressure, and $T$ is the gas
temperature. Under the assumption that they are perfectly coupled, gas
and dust share the same velocity $\mathbf{v}$, and the dust-to-gas
ratio is conserved in a Lagrangian sense.  The background potential is
provided by the central star of mass $M$: $\Phi= -GM/\sqrt{r^2+z^2}$,
where $r$ is the cylindrical radius and $z$ is the vertical distance
above the disk midplane.  There are five equations for the five flow
variables $\mathbf{v}$, $\rhog$, $\rhod$, $P$, and $T$.  The
thermodynamic constants include the specific heat $C_{\rm V} =
\Re/(\gamma-1)$ at constant volume, the ideal gas constant $\Re =
C_{\rm P}-C_{\rm V}$, the specific heat $C_{\rm P}$ at constant
pressure, and $\gamma = C_{\rm P}/C_{\rm V}$. Equation
(\ref{eqn:adiabat}) is equivalent to the condition that the flow be
isentropic [$d(P\rho_{\rm g}^{-\gamma})/dt = 0$]. 
The code which
solves the fluid equations actually employs an artificial
hyperviscosity to damp away the smallest scale perturbations
(\S\ref{ssec:code}); in writing down equations
(\ref{eqn:momentum})--(\ref{eqn:idealeos}), we have omitted the
hyperviscosity terms for simplicity.

We move to a frame co-rotating with dust-free gas at some fiducial
radius $r = R$. This frame has angular frequency $\Omega_{\rm F}$
given by (\ref{eqn:omegaf}) with 
$\OmegaK = (GM/R^3)^{1/2}$. We define a velocity $v_{\rm max}$
using the pressure support parameter
$\eta$, as given by (\ref{eqn:eta}):
\begin{equation}
\label{eqn:vmaxeta}
v_{\rm max} \equiv \left.\eta\right|_{r=R} \OmegaK R.
\end{equation}
The velocity $\vmax$ is the difference in azimuthal velocity between
a strictly Keplerian flow and dust-free gas; it is the maximum
possible difference in velocity, attained at large $\mu_0$, between
gas at the midplane and gas at altitude.
The quantities $\vmax$, $\eta$,
and the background radial pressure gradient are equivalent;
specifying one specifies the other two.
Our numerical models are labeled by $\vmax$.

In addition to moving into a rotating frame, we also replace the usual
cylindrical coordinates $(r,\phi,z)$ with local Cartesian coordinates $x=
r-R$, $y = (\phi -\Omega_{\rm F} t)R$, and $z$.\footnote{Throughout this
paper we alternate freely between subscripts $(x,y,z)$ and $(r,\phi,z)$.}
Keeping terms to
first order in $|x| \sim |z| \sim \eta R$ (see the discussion
surrounding equation \ref{eqn:deltaz}) and dropping curvature
terms, the momentum equation (\ref{eqn:momentum}) reads
\begin{equation}
\frac{d\mathbf{v}}{dt} = -2\Omega_{\rm K} \hat{\mathbf{z}} \times \mathbf{v} + 3 \Omega_{\rm K}^2 x \hat{\mathbf{x}} - \Omega_{\rm K}^2z \hat{\mathbf{z}} - \frac{1}{\rhod+\rhog} \nabla P - 2 \OmegaK^2 \eta R \hat{\mathbf{x}} \label{eqn:momentum_ss}
\end{equation}
\noindent where $d/dt = \partial / \partial t + v_i \partial /\partial x_i$ ($i
= x,y,z$). On the right-hand side, the first term is the Coriolis
acceleration, the second combines centrifugal and radial
gravitational accelerations, the third represents the vertical
gravitational acceleration from the star, and the last term arises from the centrifugal
acceleration in a frame rotating at $\Omega_{\rm F} \neq \OmegaK$.
The remaining fluid equations appear the same as
(\ref{eqn:gascontinuity})--(\ref{eqn:idealeos}), except that
$\mathbf{v}$ is now measured in a (rigidly) rotating frame.

We measure all flow variables relative to a time-independent 
reference state (subscripted ``ref''):
$$\mathbf{v} = \mathbf{v}_{\rm ref} + \w{\mathbf{v}} = \w{\mathbf{v}}$$
$$P = P_{\rm ref} + \w{P}$$
$$\rhog = \rho_{\rm g,ref} + \w{\rho}_{\rm g}$$
$$T = T_{\rm ref} + \w{T}$$
$$\rhod = \rhodzero + \w{\rho}_{\rm d} = \w{\rho}_{\rm d} \,.$$
The reference state is defined as follows. It is dust-free ($\rhodzero=0$)
and has constant gas temperature $T_{\rm ref}$.
The gas in the reference state does not shear, either
in the radial or vertical directions, but rotates 
with a fixed angular frequency $\Omega_{\rm F}$ in the inertial frame
(hence $\mathbf{v}_{\rm ref} = 0$ in the rotating frame).
In the reference state there exists a radial pressure gradient
directed outward
\begin{equation}
\label{eqn:dpdrref}
-\frac{1}{\rhoref} \frac{\p P_{\rm ref}}{\p r} = 2 \OmegaK^2 \eta R = 2\OmegaK \vmax 
\end{equation}
and a vertical pressure gradient balanced by vertical tidal gravity
\begin{equation}
\label{eqn:dpdzref}
-\frac{1}{\rhoref} \frac{\p P_{\rm ref}}{\p z} = \OmegaK^2 z\,.
\end{equation}
Equation (\ref{eqn:dpdzref}) together with equation (\ref{eqn:idealeos}) and the assumption of constant $T_{\rm ref}$ implies that the reference gas density $\rhoref$ and pressure $P_{\rm ref}$ have
Gaussian vertical distributions in $z$
with scale height $H_{\rm g} = \sqrt{\Re T_{\rm ref}} / \OmegaK$.
For simplicity we neglect the radial density gradient 
($\partial \rhoref/\partial r = 0$), as did B09.
This reference state should not be confused with our equilibrium
states of interest (\S\ref{ssec:initialconditions}),
which do shear and which do contain
dust. The reference state merely serves as a fiducial.

The flows of interest are subsonic. Mach numbers $\epsilon \equiv
\w{v}/\cs$ peak at $v_{\rm max}/\cs \sim \cs/(\OmegaK R) \sim 0.02$
for gas sound speeds
$\cs \sim 1$ km/s at $R \sim 1$ AU. Such flow is nearly incompressible:
$|\w{\rhog}|/\rhoref \sim |\w{P}|/P_{\rm ref} \sim |\w{T}| / T_{\rm ref} \sim
\epsilon^2$.  
Invoking the anelastic approximation, we keep only terms
leading in $\epsilon$ in any given equation. Equations
(\ref{eqn:gascontinuity}), (\ref{eqn:noslip}), and
(\ref{eqn:idealeos}) reduce to:
\begin{eqnarray}
\frac{d\rhog}{dt} + \rhog \nabla \cdot \mathbf{v} = \frac{\p \rhog}{\p t} + \nabla \cdot (\rhog \mathbf{v}) \approx \nabla \cdot (\rhoref \mathbf{v}) &=& 0 \, \nonumber \\ 
\label{eqn:gascontinuity_an}\\
\frac{d(\rhod/\rhog)}{dt} \approx \frac{d(\w{\rhod}/\rhoref)}{dt} \equiv \frac{d\mu}{dt} = 0 \label{eqn:noslip_an}\\
\frac{\w{P}}{\rhoref} \equiv \w{h} = \frac{\w{\rhog}}{\rhoref} \Re
T_{\rm ref} + \Re \w{T} \label{eqn:idealeos_an}
\end{eqnarray}
\noindent where we define $\mu \equiv \w{\rho}_{\rm d}/\rhoref =
\rhod/\rhoref$ and the pressure-like enthalpy $\w{h} \equiv
\w{P}/\rhoref$, and henceforth for convenience 
drop all tildes on $\rhod$, $\mu$, and
$\mathbf{v}$ (but not the other variables related to gas).  The
rightmost equalities of (\ref{eqn:gascontinuity_an}),
(\ref{eqn:noslip_an}), and (\ref{eqn:idealeos_an}) match equations
(12b), (12c), and (12e) of B09. The anelastic approximation has been
employed in the study of atmospheric convection
\citep{oguraphillips62,gough69}, stars \citep{gilmanglatzmaier81}, and
vortices in protoplanetary disks
\citep{barrancomarcus00,barrancomarcus05,barrancomarcus06}.
By eliminating the time derivative in the continuity equation
(\ref{eqn:gascontinuity_an}), we effectively ``sound-proof'' the
fluid. The simulation timestep is not limited by the sound-crossing
time but rather by the longer advection time.

We rewrite our energy equation (\ref{eqn:adiabat}) as follows: replace
$-\nabla \cdot \mathbf{v}$ 
with $d\ln \rhog/dt = -d\ln T/dt + d\ln P/dt$ to find that
\begin{eqnarray}
C_{\rm P} \frac{d\w{T}}{dt} & = & \frac{1}{\rhog} \frac{dP}{dt} \nonumber \\
 & \approx & \frac{1}{\rhoref} {\mathbf{v}} \cdot \nabla P_{\rm ref} \nonumber \\
 & \approx & - {\mathbf{v}} \cdot \left(2 \OmegaK^2 \eta R \mathbf{\hat{x}} + \OmegaK^2 z \mathbf{\hat{z}}\right) \label{eqn:adiabat_an}
\end{eqnarray}
\noindent 
where for the second line we dropped $d\w{P}/dt$ in comparison to ${\mathbf{v}}
\cdot \nabla P_{\rm ref}$, and for the third line we replaced $\rhoref^{-1}
\nabla P_{\rm ref}$ using (\ref{eqn:dpdrref}) and (\ref{eqn:dpdzref}). Equation (\ref{eqn:adiabat_an}) matches
(12d) of B09 except that for the right-hand side he has a
coefficient equal to $1 + \w{T} / T_{\rm ref}$, which we have set to
unity.

Finally, to recover the form of the momentum equation (12a) of B09,
first consider the pressure acceleration and
isolate the contribution from dust-free gas ($-\rhog^{-1} \nabla P$):
\begin{eqnarray}
-\frac{1}{\rhod+\rhog} \nabla P & = & -\left( \frac{1}{\rhod+\rhog} - \frac{1}{\rhog} \right) \nabla P - \frac{1}{\rhog} \nabla P \nonumber \\
 & \approx & \frac{\mu}{\mu+1} \left(\frac{1}{\rhog} \nabla P\right) - \frac{1}{\rhog} \nabla P \,. \label{eqn:isolate}
\end{eqnarray}
Now expand 
\begin{eqnarray}
\frac{1}{\rho_g}\nabla P & \approx & \frac{1}{\rhoref}\nabla P_{\rm ref} + \frac{1}{\rhoref}\nabla \w{P} - \frac{\w{\rho_{\rm g}}}{\rhoref^2}\nabla P_{\rm ref} \nonumber \\
 & \approx & \frac{1}{\rhoref}\nabla P_{\rm ref} + \nabla \w{h} + \frac{\w{T}}{T_{\rm ref}}\left(\frac{\nabla P_{\rm ref}}{\rhoref}\right)\nonumber  \\
 & \approx & -\left(1+\frac{\w{T}}{T_{\rm ref}}\right)(2\OmegaK^2 \eta R \mathbf{\hat{x}} + \OmegaK^2 z \mathbf{\hat{z}}) + \nabla \w{h}  \nonumber \\ 
\label{eqn:napbfinal}
\end{eqnarray}
where for the last line we used (\ref{eqn:dpdrref}) and (\ref{eqn:dpdzref}). Insertion of (\ref{eqn:isolate}) and (\ref{eqn:napbfinal}) into (\ref{eqn:momentum_ss}) yields the anelastic momentum equation (12a) of B09:
\begin{eqnarray}
\label{eqn:momentumanelastic}
\frac{d\mathbf{v}}{dt} &=& -2\OmegaK \mathbf{\hat{z}\times v} + 3\OmegaK^2 x \mathbf{\hat{x}} + \frac{\w{T}}{T_{\rm ref}} (2\OmegaK^2 \eta R \mathbf{\hat{x}} + \OmegaK^2 z \mathbf{\hat{z}}) -\nabla \w{h} \nonumber \\
&&  - \frac{\mu}{\mu+1} \left[\left(1+\frac{\w{T}}{T_{\rm ref}}\right)(2\OmegaK^2 \eta R \mathbf{\hat{x}} + \OmegaK^2 z \mathbf{\hat{z}}) - \nabla \w{h}  \right]. \nonumber \\ 
\end{eqnarray}
which isolates the driving term due to dust.

\subsection{Initial Conditions}
\label{ssec:initialconditions}

Equilibrium initial conditions (superscripted ``$\dagger$'')
are specified by five functions:
$\mu = \mu^\dagger$, 
$\w{T} = \w{T}^\dagger$, 
$\w{h} = \w{h}^\dagger$,
$\w{\rhog} = \w{\rhog}^\dagger$,
and
${\mathbf{v}} ={\mathbf{v}}^\dagger$.
For $\mu^\dagger$, we use flows characterized by a globally constant
Richardson number \citep{sekiya98,youdinshu02,chiang08}.
The conditions $Ri = $ constant, $\partial \rhog / \partial z
\ll \partial \rhod / \partial z$, and $g = -\OmegaK^2 z$ (no self-gravity) yield 
\begin{equation}
\label{eqn:mu}
\mu^\dagger(z) = \left[\frac{1}{1/(1+\mu_0)^2 + (z/z_{\rm d})^2}\right]^{1/2} - 1,
\end{equation}
where $\mu_0$ is the initial midplane dust-to-gas ratio
and
\begin{equation}
\label{eqn:dustheight}
z_{\rm d} \equiv \frac{Ri^{1/2}\,\vmax}{\OmegaK} 
\end{equation}
is a characteristic dust height. 
The dust density peaks at the midplane and decreases to zero at
\begin{equation}
\label{eqn:dustmax}
z = \pm z_{\rm max} = \pm \frac{\sqrt{\mu_0(2+\mu_0)}}{1+\mu_0} z_{\rm d}
\end{equation}
which is consistent with our order-of-magnitude expression (\ref{eqn:deltaz}).
Neither equation (\ref{eqn:mu}) nor the code accounts for self-gravity
and therefore we are restricted to modeling flows whose densities are
less than that required for the Toomre parameter of the subdisk to
equal unity (CY10; see also \S\ref{sec:discussion}). For the
minimum-mass disk of CY10, this restriction is equivalent to $\mu
\lesssim 30$.  Input model parameters include $\mu_0$, $Ri$, and
$v_{\rm max}$.

For the gas, we assume
\begin{equation}
\w{T}^\dagger = 0
\end{equation}
(initially isothermal) and solve vertical hydrostatic equilibrium for
$\w{h}^\dagger$ (the $z$-component of equation \ref{eqn:momentumanelastic}):
\begin{equation}
\label{eqn:enthalpy}
\frac{\p \w{h}^\dagger}{\p z} =  -\mu^\dagger \OmegaK^2 z \,.
\end{equation}
\noindent The functional form for $\w{h}^\dagger (z)$ is not especially
revealing and so we do not write it out here. For simplicity we assume that
$\w{h}^\dagger$ does not depend on $x$.
From $\w{h}^\dagger$ and $\w{T}^\dagger=0$
it follows from (\ref{eqn:idealeos_an}) that
\begin{equation}
\w{\rhog}^\dagger =
\frac{\rhoref \w{h}^\dagger}{ \Re T_{\rm ref}} \,.
\end{equation}
The fractional deviations
$\w{\rhog}^\dagger/\rhoref$ and $\w{P}^\dagger/P_{\rm ref}$ from the
reference state are very small, of order $\mu^\dagger (v_{\rm
  max}/\cs)^2Ri$.

It remains to specify ${\mathbf{v}}^\dagger$. Using the conditions on
$\w{h}^\dagger$ stated above, we solve for the equilibrium
(steady-state) solution to equation (\ref{eqn:momentumanelastic}):
\begin{eqnarray}
{v}_x^\dagger & = & {v}_z^\dagger  =  0 \nonumber \\
{v}_y^\dagger & = & -\frac{3}{2}\OmegaK x +
\left[\frac{\mu^\dagger(z)}{\mu^\dagger(z)+1}\right] \vmax \, . \label{eqn:vy}
\end{eqnarray}
In our reference frame rotating with the velocity of dust-free gas at $R$,
the first term on the right side of (\ref{eqn:vy}) accounts for the standard Kepler shear, 
while the second term describes how dust, which adds to inertia but
not pressure, speeds up the gas.

To $\mu^\dagger$ we add random perturbations
\begin{equation}
\label{eqn:perturbations}
\Delta\mu(x,y,z) = A(x,y) \mu^\dagger(z) [ \cos(\pi z/2z_d) + \sin(\pi z/2z_d)] \, .
\end{equation}
The amplitude $A(x,y$) is constructed in Fourier space so
that each Fourier mode has a random phase and an amplitude inversely
proportional to the horizontal wavenumber: $\hat{A} \propto
k^{-1}_\perp = (k^2_x + k^2_y)^{-1/2}$.
Because our box sizes are scaled to $z_{\rm max}$, our Fourier
noise amplitudes are largest on scales comparable to the dust layer thickness.
Thus those modes which are most likely to overturn the layer are given the
greatest initial power.
The perturbations are also chosen to be antisymmetric
about the $x$-axis so that no extra energy is injected into the system.
We take the root-mean-squared
amplitude $\Arms \equiv \langle A^2 \rangle^{1/2}$ of the perturbations to be
$10^{-4}$ or $10^{-3}$.

In summary, three input parameters $\mu_0$, $Ri$, and $\vmax$
determine our isothermal equilibrium initial conditions
(equations \ref{eqn:mu}, \ref{eqn:enthalpy}, and \ref{eqn:vy}).\footnote{While our initial conditions are isothermal, the temperature of the flow can change because of adiabatic compression/expansion and because our artificial hyperviscosity dissipates the highest wavenumber disturbances. These temperature changes are fractionally tiny because the flow is highly subsonic.}
The equilibrium solution for $\mu(z)$ is then perturbed (equation
\ref{eqn:perturbations}) by a root-mean-squared fractional amount
$\Arms$. The parameters of primary interest are $\mu_0$ and $Ri$.  
For the remaining parameters $\vmax$ and $\Arms$ we consider
three possible combinations: $(\vmax, \Arms) =
(0.025\cs,10^{-4})$ for our standard runs; $(0.025\cs,10^{-3})$
to probe larger initial perturbations;
and $(0.05\cs,10^{-4})$ to assess the effect of a stronger
radial pressure gradient.

Note that specifying $\mu_0$ and $Ri$ (and $\vmax$, though this last
variable is fixed for all of our standard runs) specifies the entire
dust and gas vertical profiles, $\rhod(z)$ and $\rhog(z)$, and by
extension the bulk height-integrated metallicity, $\Sigmad/\Sigmag
\equiv \int \rhod dz / \int \rhog dz$. We do not give an explicit
expression for $\Sigmad/\Sigmag$ because it is cumbersome
and not particularly revealing. The bulk metallicity is in some
sense the most natural independent variable because its value is given
by the background disk (for ways in which the bulk metallicity
may change, e.g., by radial particle drifts, see CY10).  We will plot
our results in the space of $\mu_0$, $Ri$, and $\Sigmad/\Sigmag$,
keeping in mind that only two of these three variables are
independent.

\subsection{Code}
\label{ssec:code}

We use the spectral, anelastic, shearing box code developed by
\citet{barrancomarcus06} and modified by B09 to
simulate well-coupled gas and dust. The code employs
shearing periodic boundary conditions in $r$, periodic boundary
conditions in $\phi$, and closed lid boundaries in $z$; the vertical
velocity $v_z$ is required to vanish at the top and bottom of the box
($z =\pm L_z/2$).

Spectral methods approximate the solution to the fluid equations as a
linear combination of basis functions.  The basis functions describe
how the flow varies in space, and the coefficients of the functions
are determined at every timestep.  For each of the periodic
dimensions, a standard Fourier basis is used, while for the vertical
direction, Chebyshev polynomials are employed.  Whereas in $r$ and
$\phi$ grid points are evenly spaced, the use of Chebyshev polynomials
in $z$ has the effect that vertical grid points are unevenly spaced;
points are concentrated towards the top and bottom boundaries of the
box, away from the midplane where the dust layer resides. Thus to
resolve the dust layer vertically, we need to increase the number of
vertical grid points $N_z$ by an amount disproportionately large
compared to the numbers of radial and azimuthal grid points $N_r$ and
$N_\phi$. See \S\ref{ssec:bsnr} for further discussion.

Spectral codes have no inherent grid dissipation;
energy is allowed to cascade down to
the smallest resolved length scales through nonlinear interactions. To
avoid an energy ``pile-up'' at the highest wavenumbers, we dissipate
energy using an artificial hyperviscosity,
given in \S3.3.3 of \citet{barrancomarcus06}.

Simulations satisfy the Courant-Friedrichs-Lewy (CFL) condition which states
that the CFL number, defined as the code timestep divided by the shortest
advection time across a grid cell, be small. In the shearing
coordinates in which the code works, that advection time is the cell
dimension divided by the local velocity over and above the Keplerian
shear, i.e., orbital velocities are subtracted off before evaluating
the CFL number.  All simulations reported in this paper are
characterized by CFL numbers less than about 0.1.

\subsection{Box Size and Numerical Resolution}
\label{ssec:bsnr}

Our standard box dimensions are $(L_r, L_\phi, L_z) = (6.4, 12.8,
8)z_{\rm max}$ and the corresponding numbers of grid points are $(N_r,
N_\phi, N_z) = (32, 64, 128)$. By scaling our box lengths $L_i$ to
$z_{\rm max}$ and fixing the numbers of grid points $N_i$, we ensure
that each standard simulation enjoys the same resolution (measured in grid
points per physical length) regardless of $Ri$, $\mu_0$, and $v_{\rm
  max}$. The vertical extent of the dust layer between $z = \pm z_{\rm
  max}$ is resolved by 22 grid points (this is less than
$[128/(8z_{\rm max})] \times 2z_{\rm max}=32$ because the
Chebyshev-based vertical grid only sparsely samples the midplane). The
radial and azimuthal directions are resolved by 10 grid points per
$2z_{max}$ length.  We choose our resolution in the vertical direction
to be greater than that of the horizontal directions because the dust
layer has finer scale structure in $z$: the dust layer becomes
increasingly cuspy at the midplane as $\mu_0$ increases.  We prescribe
the same resolution in the radial and azimuthal directions ($L_\phi/N_\phi = L_r/N_r$);
experiments with different resolutions in $r$ and $\phi$ generated
spurious results.

Too small a box size can artificially affect the stability of the dust
layer, because a given box can only support modes having integer
numbers of wavelengths inside it.
Small boxes may be missing modes that in reality overturn the layer.
We verify that for all runs in which the dust layer overturns, 
the KH mode that most visibly disrupts the layer spans more than one azimuthal
wavelength. Typically 3--5 wavelengths are discerned across the box.

\begin{figure}
\plotone{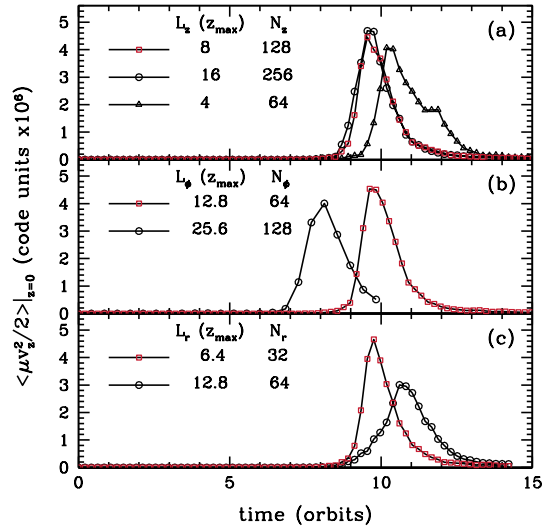}
\caption{Testing box sizes at fixed numerical resolution. For our
  standard box, $(L_r,L_\phi,L_z)=(6.4,12.8,8)z_{\rm max}$ and
  $(N_r,N_\phi,N_z)=(32,64,128)$. In each panel we vary one box dimension
  while keeping the other two dimensions fixed at their standard
  values.  In the top panel we vary $L_z$ at fixed resolution
  $N_z/L_z$. In the middle and bottom panels, $L_\phi$ and $L_r$ are
  varied in turn. All simulations in this figure have $\mu_0 = 10$,
  $Ri = 0.1$, $\vmax = 0.025 c_{\rm s}$, $\Arms = 10^{-4}$, and use
  code units $\rho_{\rm g,ref}(z=0)=\Omega_{\rm K}=H_{\rm
    g}=1$. Doubling the box dimensions from our standard values
  changes when the average vertical kinetic energy peaks by only a few
  orbits at most.  The average $\langle \rangle$ is performed over all
  $r$ and $\phi$ at fixed $z=0$.}
\label{fig:LOpt}
\end{figure}

To more thoroughly test our standard choices for $L_i$, we study how
systematic variations in
box length affect how the instability develops. For this test, we adopt a
fixed set of physical input parameters, $(Ri,\mu_0,v_{\rm
  max})=(0.1,10,0.025c_s)$, which should lead to instability
\citep{chiang08}.  
Our diagnostic is the time evolution of the
vertical kinetic energy at the midplane: $\langle \mu (t) v^2_z (t)
\rangle / 2$, where the average is over all $r$ and $\phi$ at fixed $z
= 0$ and time $t$. We vary $L_i$
and $N_i$ in tandem to maintain the same resolution from run to run,
thereby isolating the effect of box size.
Figure \ref{fig:LOpt} shows how
doubling one of the box
dimensions while fixing the other two alters
the time history of $\langle \mu v^2_z \rangle /2$.
Panel (a) demonstrates that our standard choice
for $L_z = 8 z_{\rm max}$ is sufficiently large
because the curves
for $L_z = 8 z_{\rm max}$ and $L_z = 16 z_{\rm max}$ practically overlap.
Panels (b) and (c) show that our standard choices for
$L_\phi = 12.8 z_{\rm max}$ and $L_r = 16 z_{\rm max}$ are somewhat
less adequate. The peak of the curve for
$(L_\phi,N_\phi)=(12.8z_{max},64)$ is delayed by two orbits compared
to that for $(L_\phi,N_\phi)=(25.6z_{\rm max},128)$, and the curve for
$(L_r,N_r) = (6.4z_{max},32)$ peaks an orbit earlier than that for
$(L_r,N_r) = (12.8z_{\rm max},64)$.
Nevertheless these time differences are small compared to the total time
to instability, about 10 orbits.
Moreover, the errors point in opposite directions.
Thus we expect our choices for $L_\phi$ and $L_r$ to partially
compensate for each other so that any error due to our box size in
calculating the time to instability will be less than $\sim$1 orbit.

We test how robust our results are to numerical resolution
by re-running a few simulations at twice the normal resolution
(doubling $N_i$ while fixing $L_i$). Results at high resolution
are given in \S\ref{ssec:tests}.
Every simulation is run for at least ten orbits. A typical run performed
at our standard resolution takes approximately 2.5 wall-clock hours using
56 processors on the Purdue Steele cluster. A high-resolution
run takes about 32 wall-clock hours.


\section{RESULTS}
\label{sec:results}

In our standard simulations, we fix $\vmax$ and $\Arms$ while
systematically varying $Ri$ and $\mu_0$ from run to run. Our
systematic variations of $Ri$ and $\mu_0$ correspond to systematic
variations in $\Sigmad/\Sigmag$; recall that only two of the three
parameters $Ri$, $\mu_0$, and $\Sigmad/\Sigmag$ are independent. For each
$\mu_0 \in \{ 0.3 , 1, 3, 10\}$ we adjust $Ri$ until the threshold
value $Ri_{\rm crit}$ dividing stable from unstable runs is determined
to within 0.1 dex.

Deciding by numerical simulation whether a given dust layer is stable 
or not is unavoidably subject to the finite duration
of the simulation. We define our criteria for deciding stability
in \S\ref{ssec:criteria}. Results are given in \S\ref{ssec:resultsnow}
and tested for robustness in \S\ref{ssec:tests}.

\subsection{Criteria for Stability}
\label{ssec:criteria}

Stability is assessed by two quantities: 
the midplane vertical kinetic energy
$$\langle \mu v^2_z  \rangle / 2 {\rm \,\,as \,\,a \,\,function\,\,of\,\,}t$$
where the average is performed over $r$ and $\phi$ at fixed $z=0$ and $t$, and the dust density profile
$$\langle \mu \rangle {\rm \,\,as \,\,a \,\,function\,\,of\,\,}z{\rm \,\,and\,\,}t$$
where the average is performed over $r$ and $\phi$ at fixed $z$ and $t$.  By definition, in
an ``unstable'' run, $\langle \mu v^2_z \rangle/2$ grows exponentially
over several orbital periods, and $\langle \mu \rangle$ deviates from
its initial value $\langle \mu^\dagger \rangle$ by more than 15\%.
``Stable'' simulations satisfy neither criterion.  Some runs are
``marginally unstable'' in that they satisfy the first but not the
second criterion. At the end of the standard ten-orbit duration of a
marginally unstable run, we find the kinetic energy continues to rise,
suggesting that were the run to be extended for longer than ten
orbits, the dust layer would eventually overturn.  In every instance
where we extend the duration of a marginally unstable run, we verify
that this is the case. Thus ``marginally unstable'' is practically
synonymous with ``unstable.''

\begin{figure}
\plotone{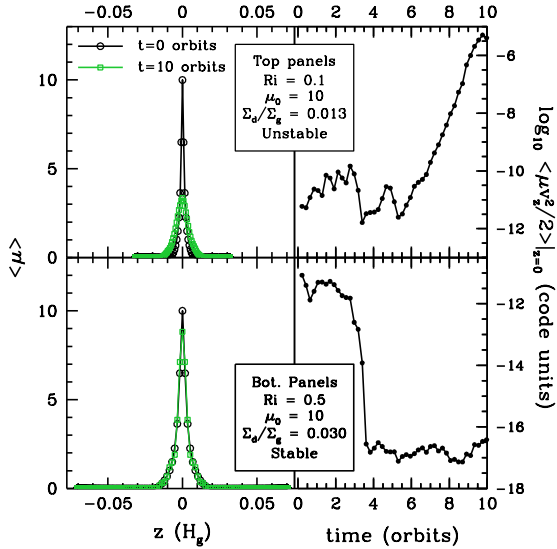}
\caption{Sample unstable (top) and stable (bottom) dust layers.  In
  the unstable case, the layer overturns and mixes dust-rich gas with
  dust-poor gas, causing the dust-to-gas ratio at the midplane to drop
  by a factor of $\sim$3 after 10 orbits (top left). As the
  instability unfolds, the vertical kinetic energy amplifies
  exponentially from $t\approx 5$--10 orbits (top right). At fixed
  $\mu_0$, the layer is stabilized by increasing the Richardson number
  or equivalently the height-integrated metallicity
  $\Sigmad/\Sigmag$. In the stable run, the dust profile changes by
  less than 15\% (bottom left) while the kinetic energy, after
  dropping precipitously, shows no indication of growing (bottom
  right).  The two runs shown use $\vmax=0.025 c_{\rm s}$ and
  $\Arms=10^{-4}$.}
\label{fig:Profile}
\end{figure}

Examples of unstable and stable runs are shown in Figure \ref{fig:Profile}. 
In the unstable simulation, after $t \approx 6$ orbits,
the kinetic energy rises exponentially. At $t \approx 9$ orbits,
the dust layer overturns and the midplane dust-to-gas ratio
falls by more than 60\%.
By contrast, in the stable simulation, after an initial adjustment period
lasting $\sim$3 orbits during which the midplane
value of $\langle \mu\rangle$ decreases by 10\%,
the kinetic energy drops by orders of magnitude to a nearly constant
value and shows no evidence of further growth.

\begin{figure}
\plotone{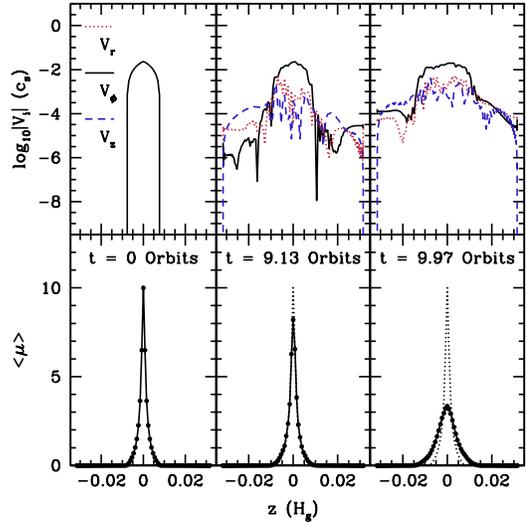}
\caption{Snapshots of absolute values of the three velocity
components (top panels) and horizontally averaged dust-to-gas ratio (bottom
panels), both as functions of height, at three instants in time.
For this unstable run,  $(Ri,\mu_0,v_{\rm max},\Arms)=(0.1,10,0.025\cs,10^{-4})$.
Velocities are taken from a grid point near the middle
of the box. The vertical shear $\partial v_\phi/\partial z$ inside
the dust layer weakens with time as dust is more uniformly mixed with
gas, and as the radial and vertical velocities grow at the expense of the
azimuthal velocity.}
\label{fig:VMuEvolve}
\end{figure}

\begin{figure}
\plotone{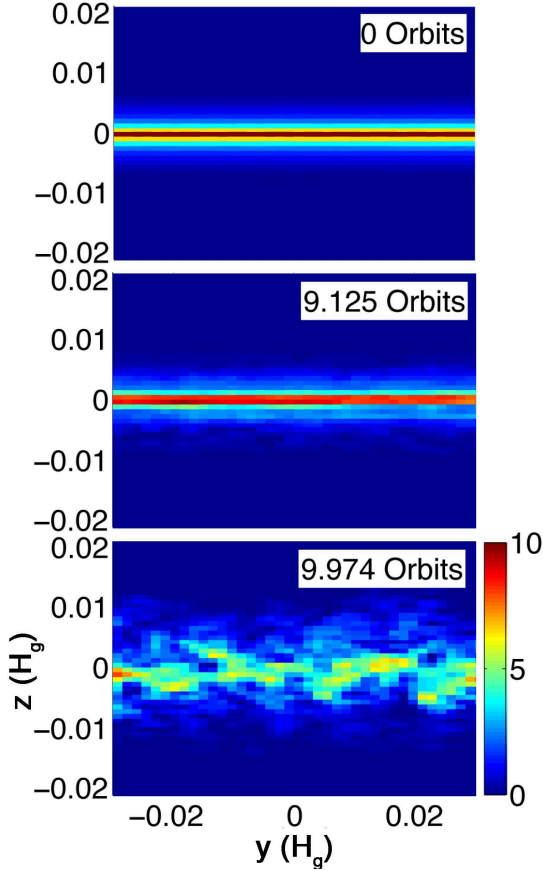}
\caption{Snapshots of $\mu (y,z)$, sampled at $r = R$ ($x=0$; the central slice
of the simulation box) for the same unstable run
shown in Figure \ref{fig:VMuEvolve}. The box
size parameters are $(L_r,L_\phi,L_z)=(0.05,0.1,0.063)H_{\rm g}$,
larger than what is shown in the figure, which zooms in for more detail.
}
\label{fig:MuColor}
\end{figure}

Figure \ref{fig:VMuEvolve} shows the evolution of $|v_i(z)|\ (i=r,\phi,z)$
and $\langle\mu(z) \rangle$ for the same unstable run of Figure
\ref{fig:Profile}. The velocity data are sampled at a single $(x,y)$
position at the center of our simulation box.
The radial and vertical velocities $|v_r|$ and $|v_z|$, initially zero, grow 
to become comparable with the shearing velocity $|v_{\phi}|$.
Figure \ref{fig:MuColor} displays corresponding snapshots of $\mu(y,z)$,
taken at a single radius $x$ near the center of our box.
Though the data in Figures \ref{fig:VMuEvolve} and \ref{fig:MuColor} 
are sampled at particular radial locations in our box,
we verify that the instability
develops similarly at all locations---as it should---unlike the ZEUS-based
simulations of \citet{chiang08}.

\subsection{Stability as a Function of $Ri$, $\mu_0$, and $\Sigmad/\Sigmag$}
\label{ssec:resultsnow}

Figure \ref{fig:M0001} maps the stable and unstable regions in $(Ri,\mu_0)$
space, for fixed $v_{\rm max} = 0.025\cs$ and $A_{\rm rms} = 10^{-4}$. 
 Figures \ref{fig:M0001RiS} and \ref{fig:M0001SM} portray the same
data using alternate but equivalent projections of parameter
space: $(Ri,\Sigmad/\Sigmag)$ and $(\mu_0,\Sigmad/\Sigmag)$, respectively. 

These plots demonstrate that there is no unique value of $Ri_{\rm
  crit}$.  Rather $Ri_{\rm crit}$ is a function of $\mu_0$, or
equivalently a function of $\Sigmad/\Sigmag$.  For bulk metallicities
$\Sigmad/\Sigmag$ near the solar value, $Ri_{\rm crit}$ is found to be
close to the classical value of $1/4$. But as $\Sigmad/\Sigmag$
decreases below the solar value, $Ri_{\rm crit}$ shrinks to $\sim$0.01
or even lower.  A least-squares fit to the four midpoints (evaluated in log space) in Figure \ref{fig:M0001}
dividing neighboring stable points (in black) and unstable points (in red or red outlined with black) yields $Ri_{\rm crit}
\propto \mu_0^{1.0}$.  This same fit, projected into metallicity
space, is shown in Figures \ref{fig:M0001RiS} and \ref{fig:M0001SM};
in metallicity space the stability boundary is not a power law.

As Figure \ref{fig:M0001SM} attests, dust-to-gas
ratios $\mu_0$ as high as $\sim$8 can be attained in disks of solar
metallicity without triggering a shear instability: see the
intersection between the dashed curve fitted to our standard
resolution data, and the dotted line representing solar
metallicity. This intersection occurs at $\mu_0 \approx 7$. Were we to
re-fit the dashed curve using the higher resolution data represented
by triangles, the intersection with solar metallicity would occur at
$\mu_0$ closer to 8.

A dust-to-gas ratio of $\mu_0 \approx 8$ is within a factor of $\sim$4 of the
Toomre threshold for gravitational fragmentation in a minimum-mass
disk (CY10; \S\ref{sec:discussion}).  We can achieve the Toomre
threshold by simply allowing for a gas disk that is $\sim$$4 \times$ more
massive than the minimum-mass nebula. Alternatively we can enrich the disk in metals
to increase $\Sigmad/\Sigmag$.
Extrapolating the boundary of stability (dashed curve) in Figure
\ref{fig:M0001SM} to higher $\Sigmad/\Sigmag$ suggests that the Toomre
threshold $\mu_0 \approx 30$ could be achieved for minimum-mass disks having
$\sim$$3\times$ the solar metallicity. The sensitivity to metallicity is also exemplified by Figure
\ref{fig:Profile}. For the same $\mu_0 = 10$, the dust layer based on
a near-solar metallicity of $\Sigmad/\Sigmag= 0.013$ overturns,
whereas one derived from a supersolar metallicity of $\Sigmad/\Sigmag=
0.030$ remains stable.

\subsection{Tests at Higher Resolution, Higher $A_{\rm rms}$, and
  Higher $v_{\rm max}$}
\label{ssec:tests}

\begin{figure}
\plotone{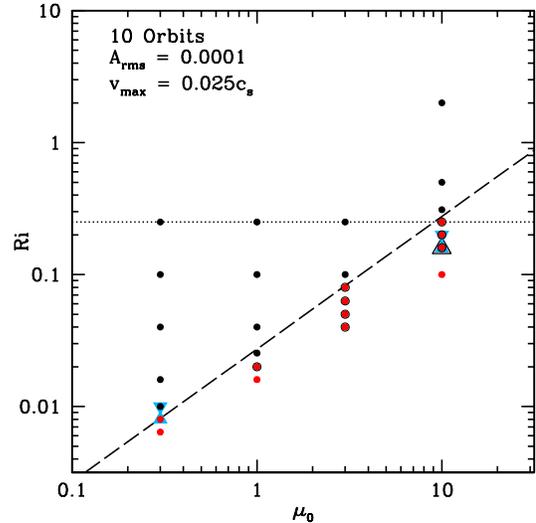}
\caption{Mapping the boundary of stability in the space of initial
  $Ri$ and $\mu_0$. Red points correspond to unstable dust layers,
  whose dust-to-gas ratios $\langle \mu \rangle$ change by more than
  15\%, and whose vertical kinetic energies grow exponentially, within
  the 10-orbit duration of the simulation.  Black points mark stable
  dust layers satisfying neither criterion. Red points outlined in
  black signify marginally unstable layers, whose kinetic energies
  rise but whose dust-to-gas ratios change by less than 15\%; these
  are essentially equivalent to red points without outlines, because
  every marginally unstable run that we extend beyond 10 orbits
  eventually becomes fully unstable. Runs performed at twice the
  standard resolution appear as triangles. Downward pointing triangles
  symbolize stable runs, upward triangles are unstable, and upward
  pointing triangles in black outline are marginally unstable.  All
  simulations use $\Arms = 10^{-4}$ and $\vmax = 0.025 \cs$.  There is
  no unique value for the critical Richardson number separating stable
  from unstable dust layers. Rather, a least-squares fit to the data
  from our standard resolution runs yields $Ri_{\rm crit} \propto
  \mu^{1.0}$, shown as a dashed line. The classical boundary $Ri_{\rm
    crit}=0.25$ is plotted as a dotted line.}
\label{fig:M0001}
\end{figure}

\begin{figure}
\plotone{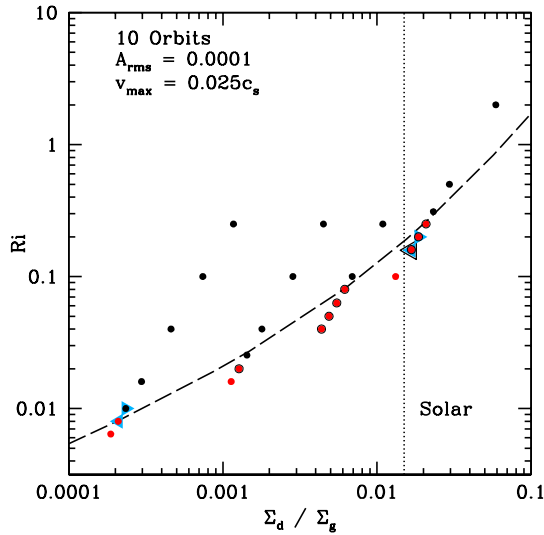}
\caption{Mapping the boundary of stability in the space of initial
  $Ri$ and bulk (height-integrated) dust-to-gas ratio
  $\Sigmad/\Sigmag$. The data are identical to those in Figure
  \ref{fig:M0001}. The labeling convention is also the same, except
  that the triangles representing high-resolution runs have adjusted
  their orientation so that they point towards the stability
  boundary. The same least-squares fit from Figure \ref{fig:M0001} is
  projected here as a dashed curve.  Solar metallicity
  $\Sigmad/\Sigmag=0.015$ \citep{lodders03} is indicated by a dotted
  line. The critical value $Ri_{\rm crit}$ dividing stable from
  unstable dusty subdisks trends with metallicity. This trend was only
  hinted at in the data of C08.}
\label{fig:M0001RiS}
\end{figure}

\begin{figure}
\plotone{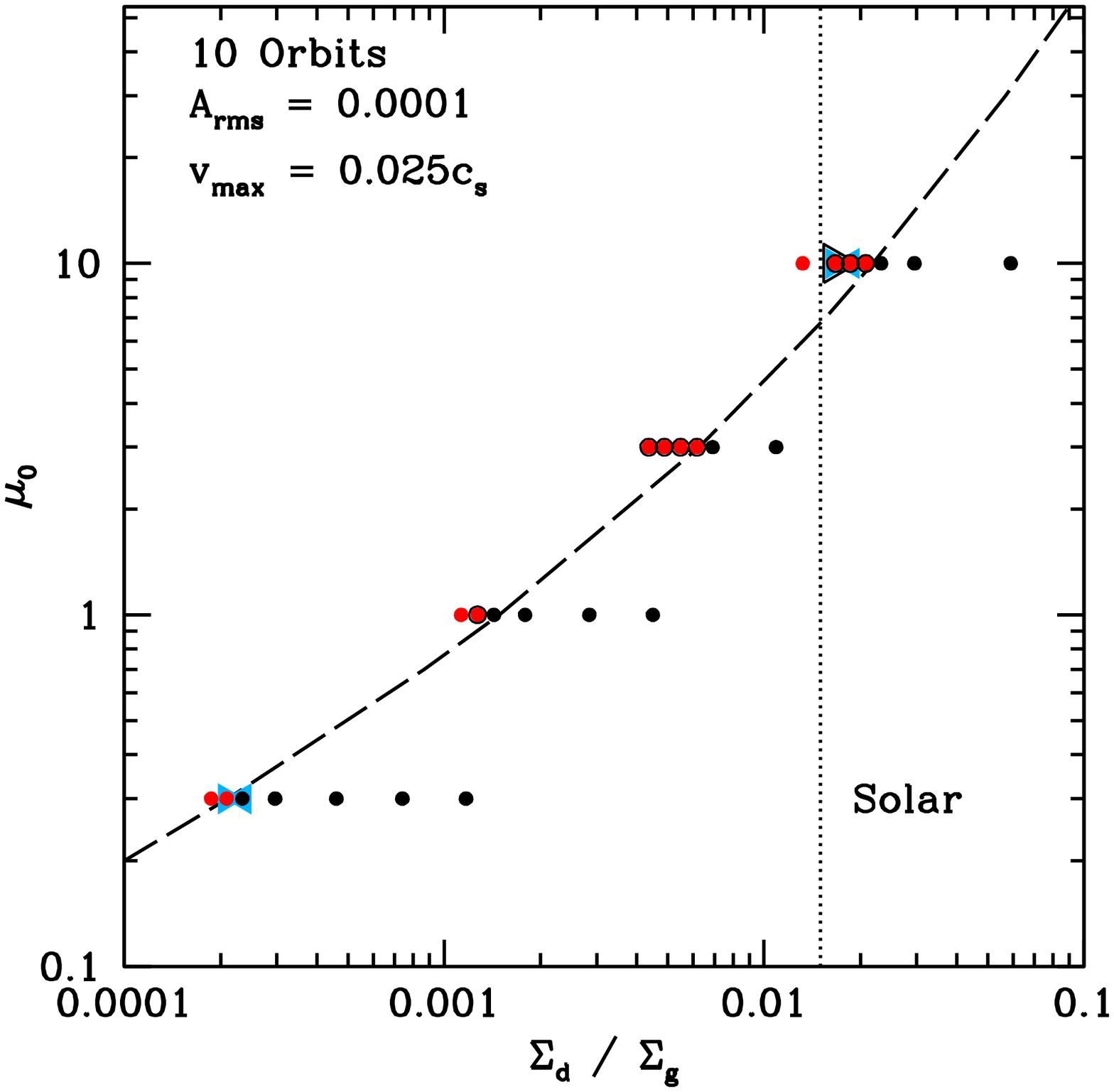}
\caption{Mapping the boundary of stability in the space of midplane
  dust-to-gas ratio $\mu_0$ and bulk (height-integrated) dust-to-gas
  ratio $\Sigmad/\Sigmag$. The data are identical to those in Figure
  \ref{fig:M0001}. The labeling convention is also the same, except
  that the triangles representing high-resolution runs have adjusted
  their orientation so that they point towards the stability
  boundary. The same least-squares fit from Figure \ref{fig:M0001} is
  projected here as a dashed
  curve. Solar metallicity $\Sigmad/\Sigmag=0.015$ \citep{lodders03}
  is indicated by a dotted line. A minimum-mass solar nebula requires
  $\mu_0 \approx 30$ for gravitational instability to ensue on a
  dynamical time (CY10). Extrapolating the boundary of stability to
 $\mu_0 \approx 30$ suggests that metallicities roughly $\sim$3 times
 solar would be required for dynamical gravitational instability in a
 minimum-mass disk. The required degree of metal enrichment would be
  proportionately less in more massive disks.}
\label{fig:M0001SM}
\end{figure}

We test how robust our determination of $Ri_{\rm crit}$ is to
numerical resolution by redoing our simulations for $\mu_0=0.3$ and
$10$ with double the number of grid points in each dimension. The
results are overlaid as blue triangles in Figures \ref{fig:M0001},
\ref{fig:M0001RiS}, and \ref{fig:M0001SM}. At $\mu_0 = 0.3$,
increasing the resolution does not change $Ri_{\rm crit}$ from its
value of 0.009.  At $\mu_0 = 10$, $Ri_{\rm crit}$ shifts downward from
0.3 to 0.2. Although we have not strictly demonstrated convergence of
our results with resolution, and although high resolution data at
other values of $\mu_0$ are missing, it seems safe to conclude that
the slope of the stability boundary in $Ri$-$\mu_0$ space is close to,
but decidedly shallower than, linear.

\begin{figure}
\plotone{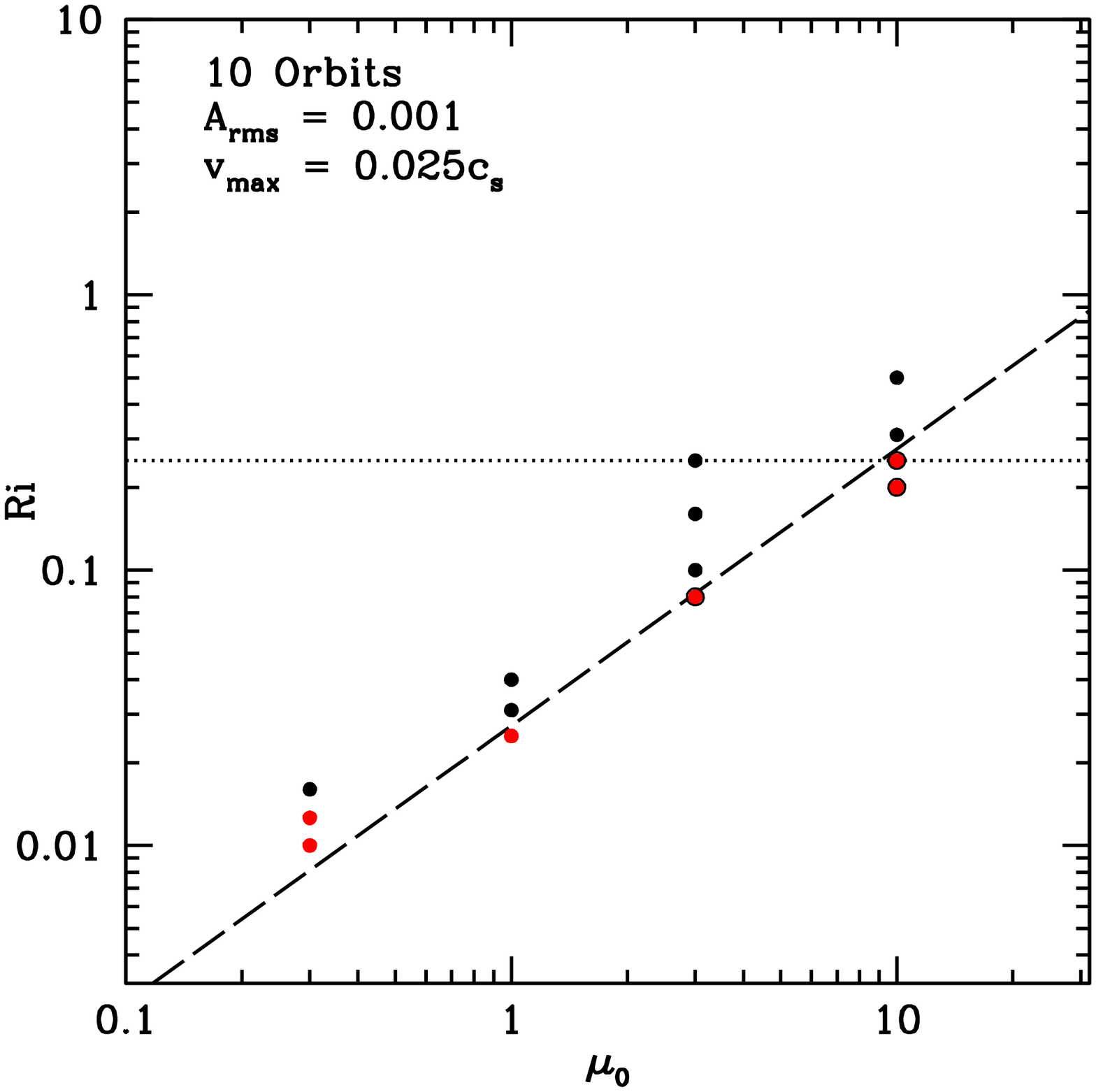}
\caption{How the stability boundary changes with stronger initial
  perturbations. This figure is the same as Figure \ref{fig:M0001}, except
  that all data correspond to $\Arms = 10^{-3}$. For comparison with
  $\Arms = 10^{-4}$, the same best-fit line of Figure \ref{fig:M0001}
  is reproduced here. Not much changes, except that $Ri_{\rm crit}$
  shifts upward by 0.2 dex at $\mu_0 = 0.3$.}
\label{fig:M001}
\end{figure}

We also test the sensitivity of our results to $\Arms$.
Increasing $\Arms$ by an order of magnitude to $10^{-3}$ shifts $Ri_{\rm crit}$
upward by $\lesssim$ 0.2 dex at $\mu_0 < 1$, but leaves
$Ri_{\rm crit}$ unchanged at larger $\mu_0$ (Figure \ref{fig:M001}).
B09 also reported some sensitivity to $\Arms$.

Tests where $\vmax$ was doubled to $0.05\cs$ reveal
no change in $Ri_{\rm crit}$ (data not shown).

\section{SUMMARY AND DISCUSSION}
\label{sec:discussion}

Where a protoplanetary disk is devoid of turbulence intrinsic to gas,
dust particles settle toward the midplane, accumulating in a sublayer
so thin and so dense that the dust-gas mixture becomes unstable. If
the first instability to manifest is self-gravitational, dust
particles are drawn further together, possibly spawning planetesimals.
If instead the layer is first rendered unstable by a
Kelvin-Helmholtz-type shearing instability (KHI), the resultant
turbulence prevents dust from settling further, pre-empting
gravitational collapse.  In this paper we investigated the conditions
which trigger the KHI, hoping to find a region of parameter space
where the KHI might be held at bay so that planetesimals can form by
self-gravity.

A fundamental assumption underlying our work is that turbulence
intrinsic to gas can, in some regions of the disk, be neglected. There
is some consensus that near disk midplanes, in a zone extending from
$\sim$1 to at least $\sim$10 AUs from the parent star, gas may be too
poorly ionized to sustain magnetohydrodynamic turbulence \citep{ilgnernelson06, baigoodman09, turneretal10}. Presumably if the magnetorotational instability \citep[e.g.,][]{balbus09} cannot operate
at the midplane, disk gas there is laminar---pending the uncertain
ability of magnetically active surface layers to stir the disk
interior \citep[e.g.,][]{turneretal10}, or the discovery of a purely
hydrodynamic form of turbulence \citep{lithwick09}. 
To get a sense of how laminar disk gas must be to permit dust sublayers
to form, Chiang \& Youdin (2010)
 compared the height to which dust particles are stirred in an
``alpha''-turbulent disk to the thickness of the sublayer
(\ref{eqn:deltaz}). They estimated that the former is smaller than the
latter when the dimensionless turbulent diffusivity $\alpha
\lesssim 3 \times 10^{-4} \OmegaK t_{\rm stop} (r/{\rm AU})^{4/7}$ for
$t_{\rm stop} < \OmegaK^{-1}$.  To place this requirement in context,
$\alpha$ values for magnetically active zones are typically quoted to
be greater than $\sim$$10^{-3}$.  Whether magnetically dead zones are
sufficiently passive for dust to settle into sublayers remains an
outstanding question.

Modulo this concern, we studied the stability of dust
layers characterized by spatially constant Richardson numbers $Ri$
using a three-dimensional, spectral, anelastic, shearing box code
\citep{barrancomarcus06} 


\noindent that models gas and dust as two perfectly
coupled fluids \citep{barranco09}.  We found that stability is not
characterized by a single critical Richardson number. Rather the value
of $Ri_{\rm crit}$ distinguishing layers that overturn from those that
do not is a nearly linear function of the midplane dust-to-gas ratio
$\mu_0$ (Figure \ref{fig:M0001}).  Dust-rich sublayers having $\mu_0
\approx 10$ have $Ri_{\rm crit} \approx 0.2$---near the canonical
value of 1/4---while dust-poor sublayers having $\mu_0 \approx 0.3$
(still orders of magnitude dustier than well-mixed gas and dust at
solar abundances) have $Ri_{\rm crit}$ as low as 0.009.

Previous studies \citep[e.g.,][]{sekiya98,youdinshu02,youdinchiang04}
assumed a universal critical Richardson
number of 1/4. This popular assumption seems correct
only for dust-rich layers having $\mu_0$ so large they are
on the verge of gravitational instability. For less dusty midplanes,
the assumption appears to be incorrect. Our numerical results
are roughly consistent with those of \citet{chiang08}, who also found
evidence that $Ri_{\rm crit}$ decreases with decreasing $\mu_0$.
Comparing his Table 2 with our Figure \ref{fig:M0001} shows that
his constraints on $Ri_{\rm crit}$ are, for the most part,
compatible with those presented here, for the range $\mu_0 \approx
0.3$--10 where our respective data overlap.
Our findings supersede those of \citet{chiang08} insofar as
we have explored parameter space more finely and systematically, at
greater and more uniform resolution, with numerical methods better
suited for subsonic flows.

Our results turn out to be consistent with the classical
Richardson criterion---which states only that $Ri < 1/4$ is necessary,
not sufficient, for instability---even though the criterion
as derived by \citet{miles61} applies only to two-dimensional
flows, which our dust layers are not. Our simulations demonstrate
that the criterion can still serve as a
useful guide for assessing stability in disks having bulk metallicities
ranging from subsolar to slightly super-solar values---with the proviso
that the actual Richardson number dividing KH-stable from KH-unstable flows,
while $< 1/4$, is generally not equal to 1/4.

Why isn't the Richardson criterion for instability
sufficient in rotating dust disks?
The criterion considers the competition between the destabilizing
vertical shear and the stabilizing influence of buoyancy, which causes
fluid parcels to oscillate about their equilibrium positions at the
Brunt-V\"ais\"al\"a frequency. However, there exists another
stabilizing influence, ignored by the Richardson number, provided by
the radial Kepler shear \citep{ishitsusekiya03}.  In the limit $\mu_0
\ll 1$, the Brunt frequency (\ref{eqn:brunt}) becomes negligible
relative to the Kepler shearing frequency (\ref{eqn:keplershear}),
suggesting stability now depends on the competition between the
destabilizing vertical shear and stabilizing radial Kepler shear. We
expect the flow to be stable as long as the Kepler shear can wind
up unstable eigenmodes to higher radial wavenumbers before their
amplitudes grow large enough to trigger nonlinear effects. This
suggests that we replace the Richardson number with a ``shearing
number,'' defined by analogy as the square of the ratio of the Kepler
shearing frequency to the vertical shearing frequency:
\begin{equation} \label{eqn:sh}
	Sh \equiv \frac{\left| \partial \Omega / \partial \ln r \right|^2}{(\p v_\phi/\p z)^2} \propto \left( \frac{\Delta z}{\Delta v_\phi} \right)^2 \propto Ri \frac{1+\mu_0}{\mu_0} 
\end{equation}
where we have used (\ref{eqn:verticalshear}) and (\ref{eqn:deltaz}).
By assuming $Sh$ is constant for marginally stable dust profiles, we
arrive at the relation
\begin{equation}
Ri_{\rm crit} \propto \mu_0 \,\,\, {\rm for}\,\,\, \mu_0 \ll 1 \,.
\end{equation}
What is surprising is that this trend, although expected to hold only
for $\mu_0 \ll 1$, appears to hold approximately
for all $\mu_0$, according to our simulation
results in Figure \ref{fig:M0001}.
For $\mu_0\gtrsim 1$, we would have expected from (\ref{eqn:sh}) that
$Ri_{\rm crit}$ asymptote to a constant; but it does not. Our higher
resolution runs do suggest the stability curve slightly flattens
at $\mu_0 \approx 10$, but such deviations seem too small to be fully
explained using
arguments relying purely on the shearing
number.  

To explain the observed trend, we might co-opt the methods
of \citet{ishitsusekiya03}, who linearized and numerically
integrated the 3D equations of motion for the dust layer.
For their particular choice of background
vertical density profile,
they solved for the maximum growth factors for the
most unstable KH modes (see also \citealt{knoblochspruit85}
who considered the axisymmetric problem). We would need
to replace their assumed profile with our profiles having
spatially constant $Ri$. Perhaps our numerically
determined stability curve $Ri_{\rm crit} (\Sigmad/\Sigmag)$
corresponds to a locus of fixed maximum growth factor.

Gravitational instability occurs on a dynamical time when the dust
layer's Toomre $Q \approx M/[2\pi r^3\rho_{\rm g}(1+\mu_0)]$ reaches
unity \citep{toomre64,goldreichbell65}. For $\rho_{\rm g}$ given by
the minimum-mass solar nebula, this occurs when $\mu_0 \approx 30$,
fairly independently of $r$ \citep{chiangyoudin10}. 
Of course in more massive gas disks (greater $\rhog$), the requirement on $\mu_0$
is proportionately lower.
Figure \ref{fig:M0001SM} shows
that for disks having bulk metallicities $\Sigmad/\Sigmag$ equal to
the solar value of 0.015, the dusty sublayer can achieve $\mu_0
\approx 8$ before it becomes KH unstable. Taken at face value, such a
marginally KH-stable subdisk, embedded in a gas disk having $30/8
\approx 4$ times the mass of the minimum-mass solar nebula, would
undergo gravitational instability on the fastest timescale imaginable,
the dynamical time.  The case that planets form from disks
several times more massive than the minimum-mass solar nebula
is plausible \citep[e.g.,][]{goldreichetal04,lissaueretal09}.

An alternate way of crossing the Toomre threshold is to allow the bulk metallicity
$\Sigmad/\Sigmag$ to increase above the solar value of 0.015.
Extrapolating the boundary of stability in Figure \ref{fig:M0001SM} to
 $\mu_0 \approx 30$ suggests that metallicities roughly $\sim$3 times
 solar would be required for dynamical gravitational instability in a
minimum-mass disk. There are several proposed ways to achieve
supersolar metallicities in some portions of the disk, among them radial
pileups \citep{youdinshu02}
or dissipative
gravitational instability
(\citealt{ward76}; \citealt{coradinietal81}; \citealt{ward00}; \citealt{youdin05a}; \citealt{youdin05b}; see also the introduction of \citealt{goodmanpindor00}).

None of the ways we have outlined 
for achieving gravitational
instability rely on the streaming instability or turbulent
concentration of particles, mechanisms that we have criticized in
\S\ref{ssec:versus}.  Nevertheless our scenarios may be too optimistic
because all our dust profiles are predicated on the assumption of a
spatially constant $Ri$. This assumption tends to generate strong
density cusps at the midplane that might not be present in reality. In
a forthcoming paper we will relax the assumption of spatially constant
$Ri$ and measure the maximum $\mu_0$ attainable, as a function of
metalllicity $\Sigmad/\Sigmag$, by simulating explicitly the settling
of dust towards the midplane.

\acknowledgments We thank Daniel Lecoanet, Eve Ostriker, Prateek
Sharma, Jim Stone, and Yanqin Wu for discussions. An anonymous referee
provided a thoughtful and encouraging report that helped to place our work in a
broader context. This research was supported by the National Science
Foundation, in part through TeraGrid resources provided by Purdue
University under grant number TG-AST090079.

\end{document}